\documentclass[aps, prl, reprint, superscriptaddress, amsmath, amssymb]{revtex4-2}

\usepackage{graphicx}
\usepackage{dcolumn}
\usepackage{bm}
\usepackage[colorlinks,urlcolor=blue, citecolor=blue,linkcolor=blue]{hyperref}
\usepackage{booktabs}
\usepackage{overpic}
\usepackage{color}

\newcommand{\ee}{e^+e^-}
\newcommand{\jpsi}{J/\psi}
\newcommand{\psip}{\psi(3686)}

\newcommand{\BESIIIorcid}[1]{\href{https://orcid.org/#1}{\hspace*{0.1em}\raisebox{-0.45ex}{\includegraphics[width=1em]{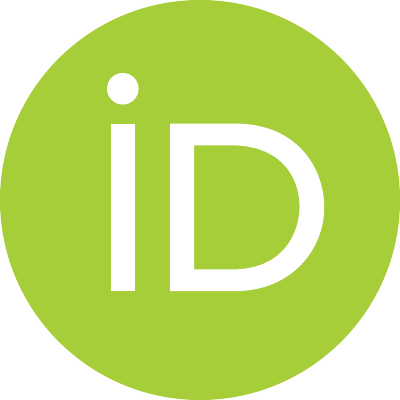}}}}

\begin{document}

\title{\boldmath Observation of a New Excited $\Sigma$ State in $\psip\to\bar{p}K^+\Sigma^0+\mathrm{c.c.}$}
\author{BESIII Collaboration}
\thanks{Full author list given at the end of the article.}
\noaffiliation

\begin{abstract}
Using a data sample of $(2712.4\pm14.3)\times10^6\ \psi(3686)$ events collected with the BESIII detector, a partial-wave analysis of the process $\psi(3686)\to\bar{p}K^+\Sigma^0+\mathrm{c.c.}$ is performed. A new excited $\Sigma$ baryon state is observed with a statistical significance of $11.9\sigma$, and the mass and width are measured as $(2334.7\pm7.9\pm16.0)\;\mathrm{MeV}/c^2$ and $(206.3\pm9.5\pm18.4)\;\mathrm{MeV}$, respectively. The spin-parity of the new state is favored to be $3/2^-$, and the branching fraction of $\psip\to\bar{\Sigma}(2330)^0\Sigma^0+\mathrm{c.c.}$ is determined to be $(4.47\pm0.58\pm1.52)\times10^{-6}$. In addition, the branching fraction of $\psi(3686)\to\bar{p}K^+\Sigma^0+\mathrm{c.c.}$ is determined to be either $(2.44\pm0.20\pm0.08)\times10^{-5}$ or $(1.73\pm0.29\pm0.06)\times10^{-5}$, taking interference with the continuum into account. The first uncertainties are statistical and the second  systematic.
\end{abstract}

\maketitle
Within the quark model, baryons, which consist of three quarks, represent the simplest system in which the three colors of quantum-chromodynamics (QCD) neutralize to form a colour singlet. Lattice QCD has played a crucial role in mapping the baryon spectrum, and studies of light-quark baryons yield insight into confinement and chiral-symmetry breaking in the non-perturbative regime. Nevertheless, despite the impressive successes of both the quark model and lattice QCD~\cite{Edwards:2011jj}, our knowledge of baryon spectroscopy remains fragmentary; many fundamental questions are still open~\cite{Klempt:2009pi}. A long-standing puzzle is the “missing-resonance” problem: far fewer excited baryons have been observed than are predicted by quark-model and lattice calculations. Consequently, more extensive experimental and theoretical investigations are required to clarify the baryon spectrum.

Experimentally, the search for predicted but still undiscovered resonances continues worldwide. Most previous experiments have concentrated on energies below $2.3\;\mathrm{GeV}$~\cite{ParticleDataGroup:2024cfk}, whereas studies of high-mass excitations have stagnated for almost five decades. This gap hampers both phenomenological analyses and further tests of QCD. Excited baryons are difficult to identify: their large natural widths (short lifetimes) and the small mass splittings between neighboring states lead to strongly overlapping signals. Partial-wave analysis (PWA) disentangles these overlapping resonances and extracts their masses, widths, spin-parity quantum numbers, and partial decay widths. Precise spectroscopic information provides an essential bridge between experiment and theory, particularly for QCD. Recent theoretical work~\cite{Menapara:2024wpb} suggests that the properties of high-mass excitations can offer critical insight into strong-interaction dynamics and the underlying quark structure of baryons. Extending PWA studies to higher masses may reveal new resonances that challenge existing models and refine our understanding of baryon spectroscopy.

At present, most information on excited hyperons composed of three light quarks comes from $\bar{K}N$ scattering~\cite{ParticleDataGroup:2024cfk}. Charmonium decays produced in $e^+e^-$ annihilation provide a complementary, clean environment for baryon-resonance studies: low background, excellent energy resolution, and large data samples~\cite{BESIII:2020nme,BESIII:2009fln,BESIII:2021cxx,BESIII:2024lks}, together with powerful PWA techniques, enable sensitive searches for rare decay modes. These advantages have been demonstrated in BESIII analyses of excited baryons in $J/\psi$ and $\psi(3686)$ decays~\cite{BESIII:2022cxi,BESIII:2022fhe,BESIII:2023syz,BESIII:2024vqu,BESIII:2024jgy}, establishing BESIII as a key facility for baryon spectroscopy.

In this Letter, we report a PWA of the decay $\psi(3686)\to\bar{p}K^+\Sigma^0$ based on $(2712.4\pm14.3)\times10^6\ \psi(3686)$ events collected with the BESIII detector. A new excited $\Sigma$ state is observed for the first time with a statistical significance of $11.9\sigma$; its spin-parity is preferred to be $3/2^-$ and its branching fraction is determined. Charge-conjugate modes are implied throughout unless stated otherwise.

The BESIII detector~\cite{BESIII:2009fln} records symmetric $e^+e^-$ collisions provided by the BEPCII storage ring~\cite{Yu:2016cof} at center-of-mass energies between $1.84$ and $4.95\;\mathrm{GeV}$, which is described in detail in Refs.~\cite{BESIII:2009fln,Yu:2016cof,BESIII:2020nme,Zhang:2022bdc}. The decay $\psip\to\bar{p}K^+\Sigma^0$ is reconstructed via the decay chain $\Sigma^0\to\Lambda\gamma$ and $\Lambda\to p\pi^{-}$. Monte Carlo (MC) simulated event samples produced with a {\sc geant4}-based~\cite{GEANT4:2002zbu} software package, which includes the geometric description of the BESIII detector and the detector response, are used to determine detection efficiencies and to estimate backgrounds. The simulation models the beam energy spread and initial state radiation (ISR) in the $\ee$ annihilations with the generator {\sc kkmc}~\cite{Jadach:1999vf, Jadach:2000ir}. The inclusive MC sample includes the production of the $\psip$ resonance, the ISR production of the $\jpsi$, and the continuum processes incorporated in {\sc kkmc}~\cite{Jadach:1999vf, Jadach:2000ir}. To perform the PWA for $\psip\to\bar{p}K^+\Sigma^0$, a sample of 10 million simulated events is generated with the {\sc kkmc} generator~\cite{Jadach:1999vf, Jadach:2000ir} according to the phase space (PHSP) model. The detection efficiency is determined with a sample of 1 million simulated events according to our PWA result. The subsequent decays are processed via {\sc evtgen}~\cite{Ping:2008zz, Lange:2001uf} according to the measured branching fractions provided by the Particle Data Group (PDG)~\cite{ParticleDataGroup:2024cfk}, and the remaining unmeasured decay modes are generated with {\sc lundcharm}~\cite{Chen:2000tv}.

The selection criteria for the charged tracks, photon showers, and particle identification (PID) for the proton, kaon and pion follow the previous BESIII analysis~\cite{BESIII:2012koo}. $\Lambda$ candidates are reconstructed using $p\pi^-$ combinations constrained to originate from a common vertex. To suppress background, the $p\pi^-$ invariant mass is required to lie within $7.5\;\mathrm{MeV}/c^2$ of the nominal $\Lambda$ mass. To improve the momentum and energy resolution and suppress background, a four-constraint (4C) kinematic fit is applied under the hypothesis of $\ee\to\bar{p}K^+\Lambda\gamma$, enforcing energy-momentum conservation from the initial $\ee$ to the final state. For events with more than one photon candidate, the combination with the minimum $\chi^2_\mathrm{4C}$ is retained. The resulting $\chi^2_\mathrm{4C}$ is required to be less than 45. To reject background contributions from $\psi(3686)\to\bar{p}K^+\Lambda$ and $\psi(3686)\to\bar{p}K^+\Lambda\gamma\gamma$, we impose the requirement that the $\chi^2_\mathrm{4C}$ for the signal hypothesis be smaller than those for either background hypothesis.

Background studies with the inclusive MC sample and the generic tool \textsc{TopoAna}~\cite{Zhou:2020ksj} show that the dominant sources are $\psi(3686)\to\gamma\chi_{cJ}$, $\chi_{cJ}\to\bar{p}K^+\Lambda$. These are suppressed by requiring the recoil mass of the photon to exceed the nominal $\chi_{c2}$ mass by at least $15\;\mathrm{MeV}/c^2$~\cite{ParticleDataGroup:2024cfk}. Continuum background is evaluated with $401.0\pm4.0\;\mathrm{pb}^{-1}$ of off-resonance data taken at $\sqrt{s}=3.65\;\mathrm{GeV}$~\cite{BESIII:2024lks}; the corresponding event yield is normalized to $606.8\pm106.8$ events for the $\psip$ energy point using the method of Ref.~\cite{BESIII:2022cxi}, which derives the normalization factor from off-resonance fits.

\begin{figure}[htb]
    \centering
    \includegraphics[width=0.95\linewidth]{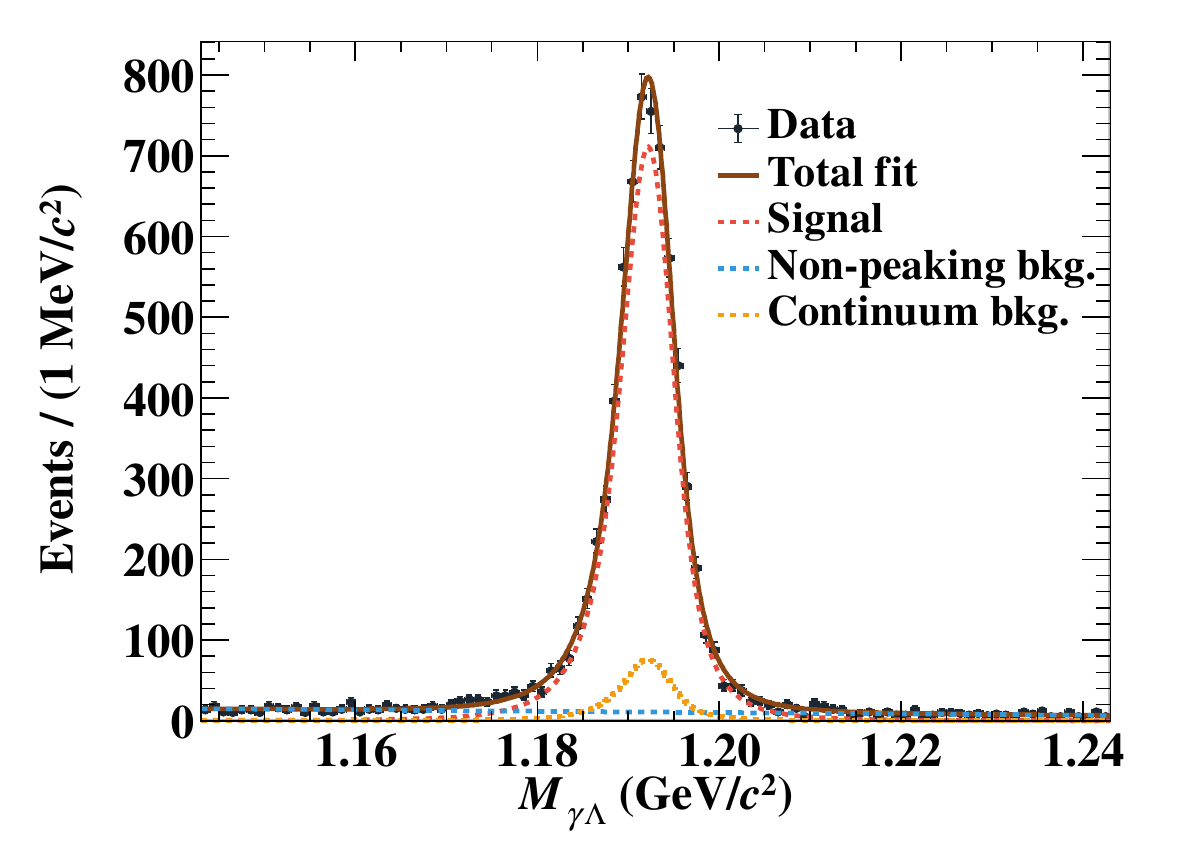}
    \caption{Fit to the $M_{\gamma\Lambda}$ distribution.  
    Dots with error bars are data; the brown solid curve shows the total fit; the red dashed curve is the signal; the blue dashed curve is the non-peaking background; the yellow dashed curve is the continuum background.}
    \label{fig:fit_m}
\end{figure}

The signal yield is determined applying an extended unbinned maximum-likelihood fit to the $M_{\gamma\Lambda}$ distribution. The signal shape is taken from MC simulation convolved with a Gaussian resolution function; the non-peaking background is parameterized by a first-order Chebyshev polynomial, while the continuum background uses the normalized shape from the $\sqrt{s}=3.65\;\mathrm{GeV}$ data with its yield fixed.   Figure~\ref{fig:fit_m} shows the fit result. The number of observed events in data is obtained to be $6099.9\pm85.5$ with an efficiency of $17.0\%$ determined from the PWA result. Following the same method as used in Ref.~\cite{BESIII:2012koo}, a branching fraction of $\psip\to\bar{p}K^+\Sigma^0$ is determined to be $(2.06\pm0.03)\times10^{-5}$, where the uncertainty is statistical only, disregarding any interference between continuum and resonant contributions.  The effect of interference between continuum and resonance amplitudes on the branching fraction is estimated with the method of Ref.~\cite{Guo:2022gkg}. The interference ratio $r_{R}^f$, which characterizes the relative effect of this interference on the branching fraction, is defined as $r_{R}^f\equiv 2AB\sin\phi/\hbar c$~\cite{Guo:2022gkg}. Specifically, the corrected branching fraction accounting for interference can be simply expressed as $\mathcal{B}=\mathcal{B}_0 (1+r_{R}^f)$, where $\mathcal{B}_0$ denotes the branching fraction without considering interference effects. Using the $\psi(3686)$ scan data~\cite{BESIII:2023euh,BESIII:2024dmr}, we obtain $A=278.033\;\mathrm{pb}^{1/2}$; the value $B=6.74\;\mathrm{GeV}/c^2$ is taken from Ref.~\cite{Guo:2022gkg}. The two phase-angle solutions, $\phi=(1.872\pm0.078)\;\mathrm{rad}$ and $(-2.102\pm0.143)\;\mathrm{rad}$, yield interference ratios of $r_{R}^f=(18.1\pm1.5)\%$ and $(-16.4\pm2.7)\%$, respectively.

The PWA employs the helicity-amplitude formalism~\cite{Richman:1984gh,Chung:1971ri} as implemented in the open-source framework \textsc{TF-PWA}~\cite{Jiang:2020tf}. The basic procedure follows the PWA workflow of Ref.~\cite{BESIII:2024jgy} for $\psip\to\Lambda\bar{\Sigma}^0\pi^0$, with no deviations except for the final-state particles. To enhance signal purity, we require $1.181<M_{\gamma\Lambda}<1.201\;\mathrm{GeV}/c^2$, leaving 6\,586 events. Mis-combination and continuum backgrounds are described using $\Sigma^0$ side-bands ($1.161$–$1.171$ and $1.211$–$1.221\;\mathrm{GeV}/c^2$) and the $\sqrt{s}=3.65\;\mathrm{GeV}$ data; their yields are estimated to be $276.0\pm16.7$ and $573.0\pm100.6$, respectively, giving a signal purity of $87.0\%$. Decay amplitudes are constructed from sequential helicity amplitudes and relativistic Breit–Wigner propagators for intermediate resonances; non-resonant ($NR$) terms are set to unity. Blatt–Weisskopf barrier factors~\cite{Chung:1993da} are included for resonant but not for the $NR$ contributions.

\begin{figure}[htb]
    \centering
    \begin{overpic}[width=0.95\linewidth]{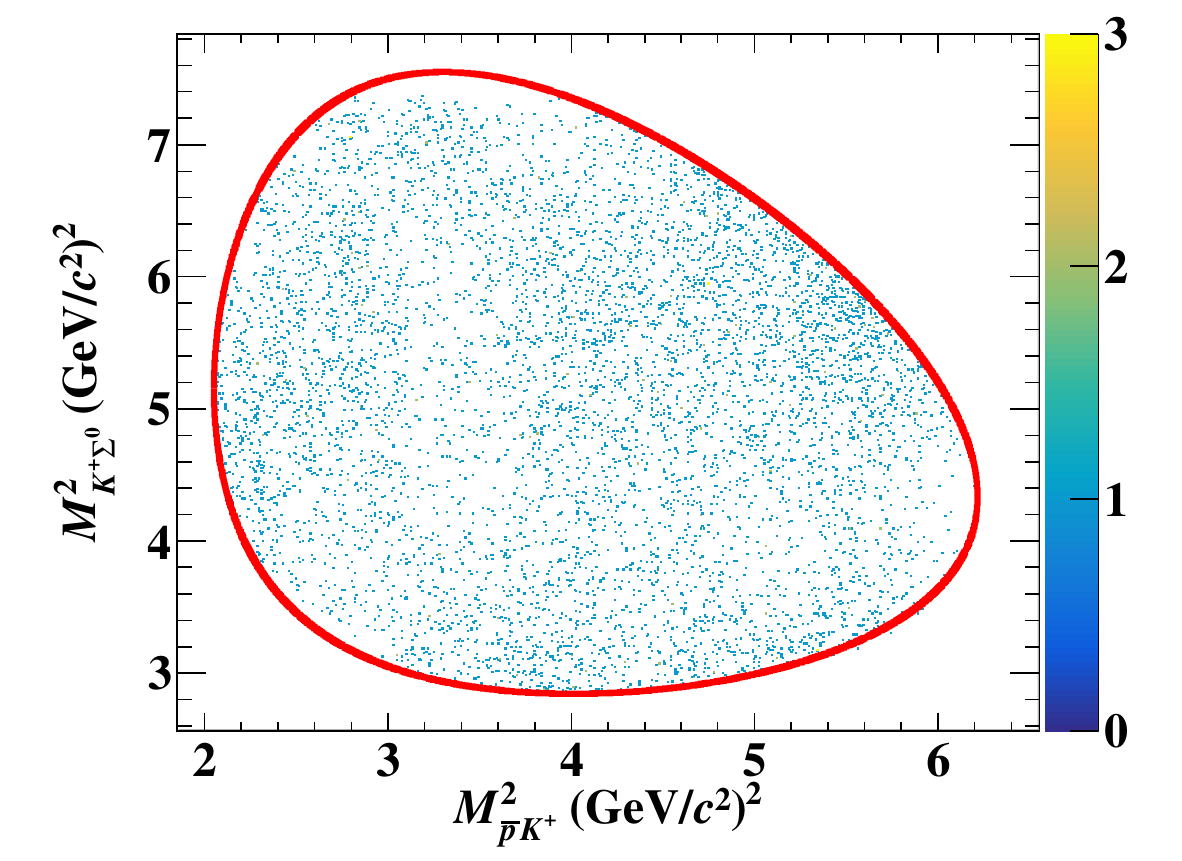}
        \put(75,60){(a)}
    \end{overpic}
    \begin{overpic}[width=0.95\linewidth]{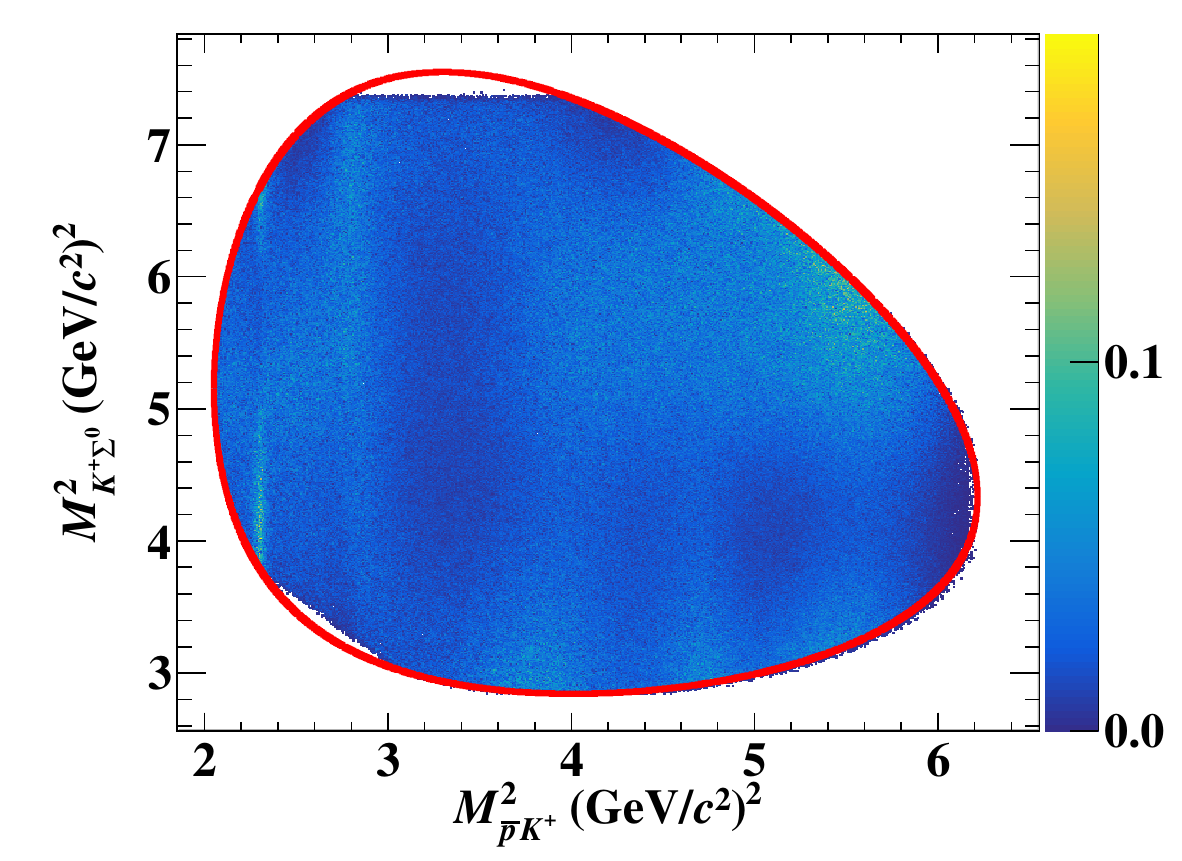}
        \put(75,60){(b)}
    \end{overpic}
    \caption{Dalitz plots of $M^2_{\bar{p}K^+}$ versus $M^2_{K^+\Sigma^0}$ of (a) data and (b) the normalized PWA fit result, and the red curves correspond the kinematic boundary.}
    \label{fig:pwa_dalitz}
\end{figure}

\begin{figure*}[htb]
    \centering
    \includegraphics[width=0.27\textwidth]{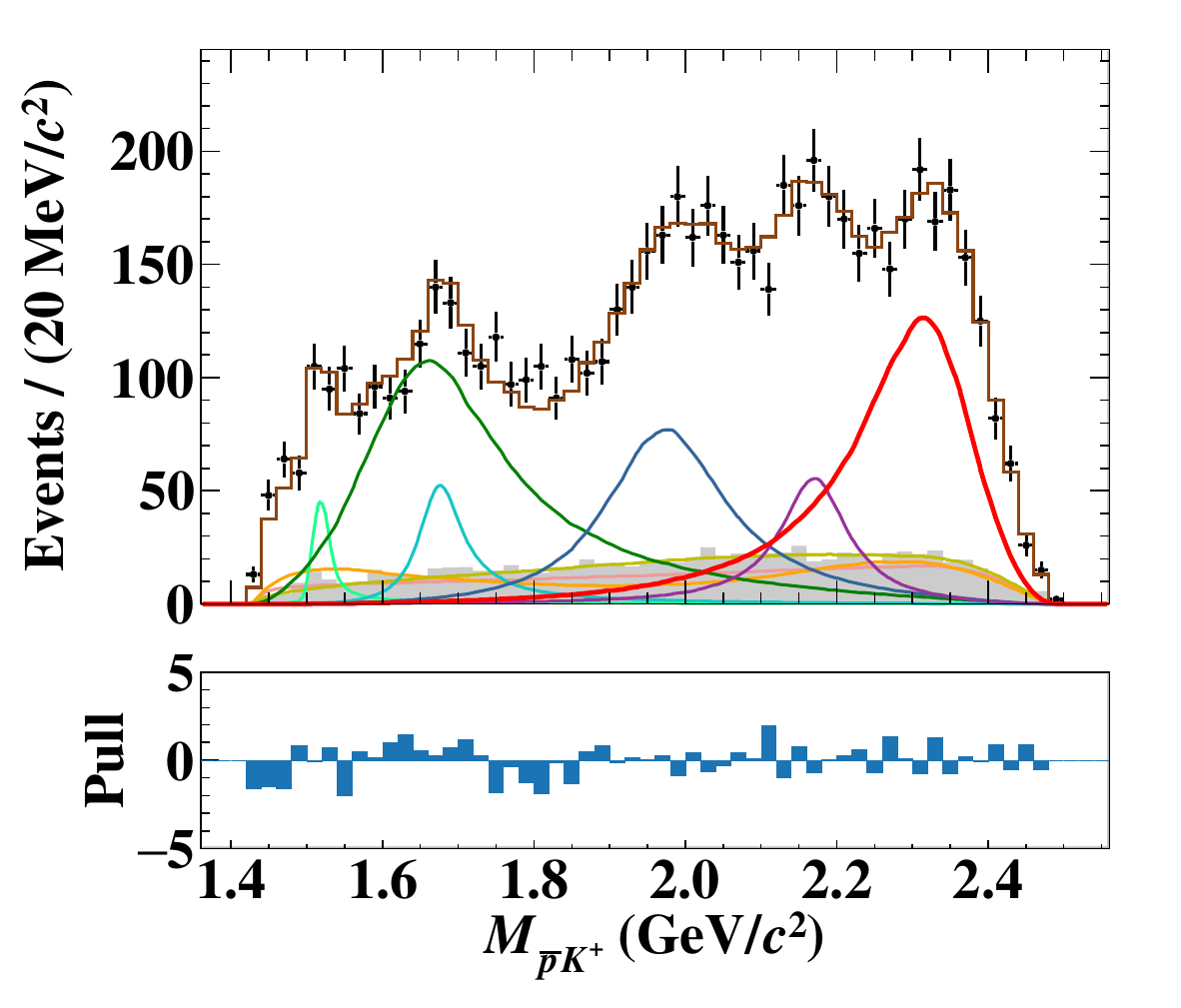}
    \includegraphics[width=0.27\textwidth]{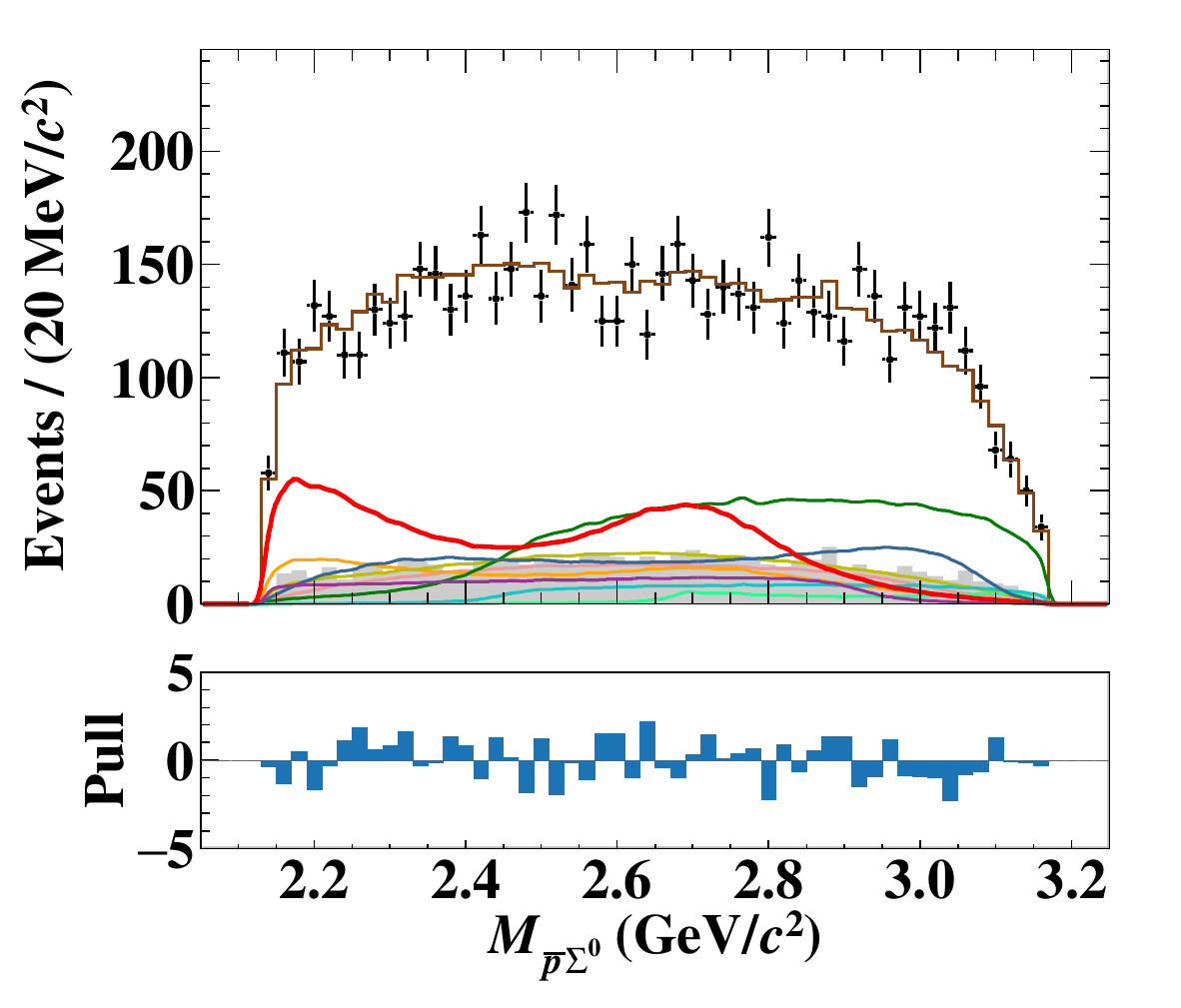}
    \includegraphics[width=0.27\textwidth]{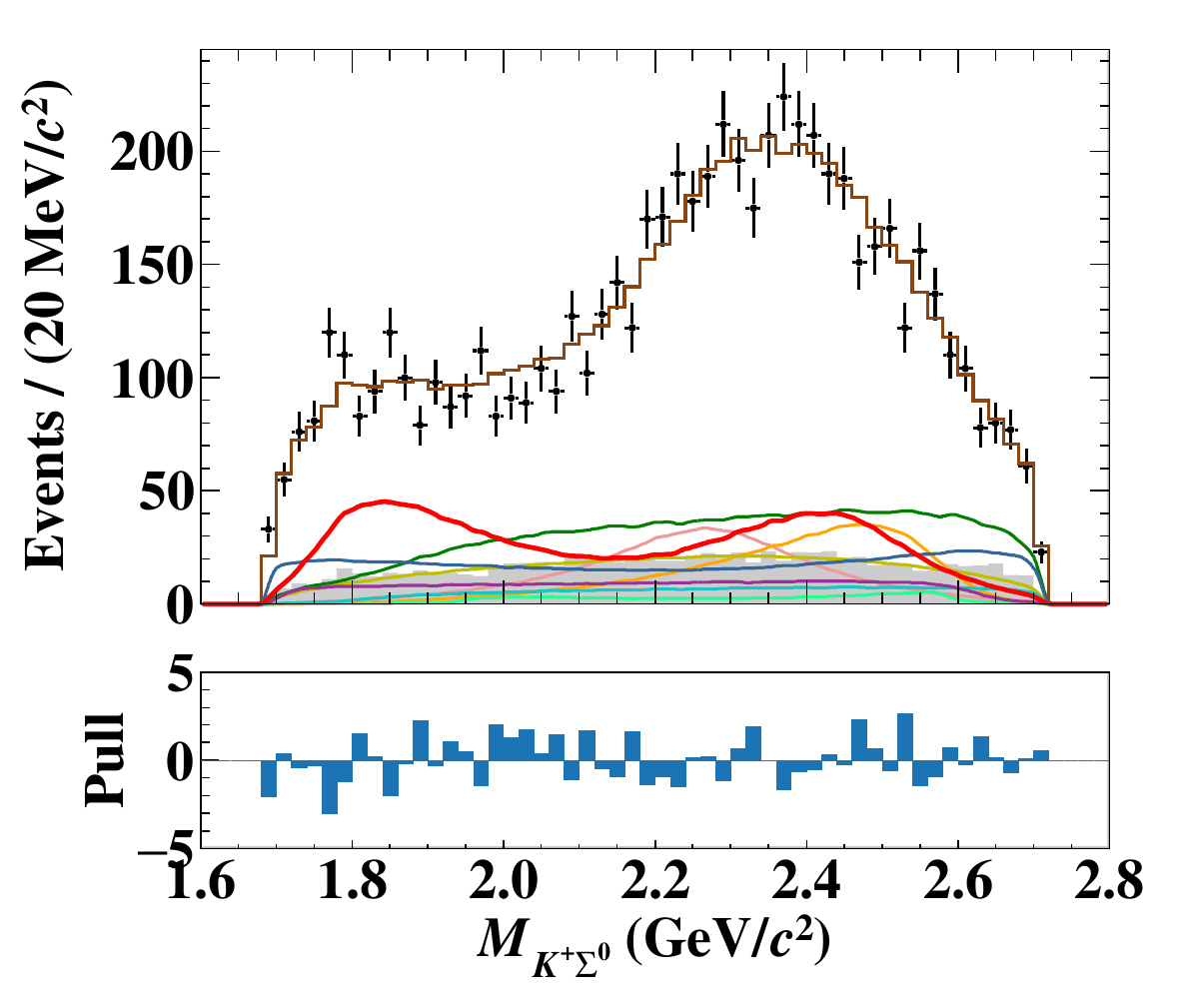}
    \includegraphics[clip, trim=2cm 4cm 10cm 0cm,width=0.135\textwidth]{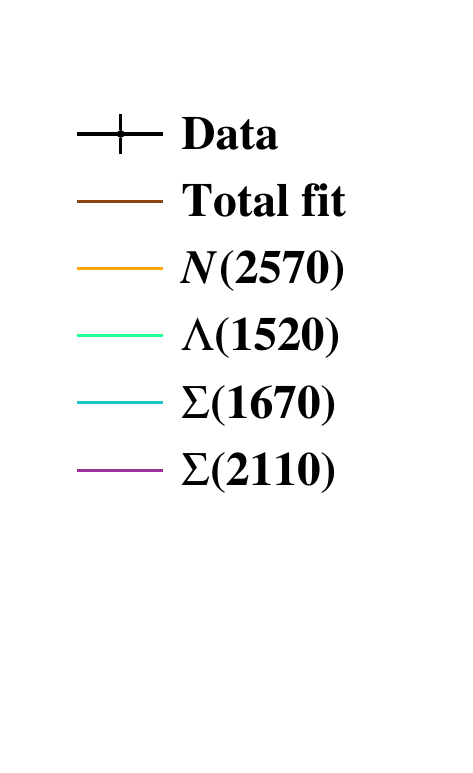}
    \includegraphics[width=0.27\textwidth]{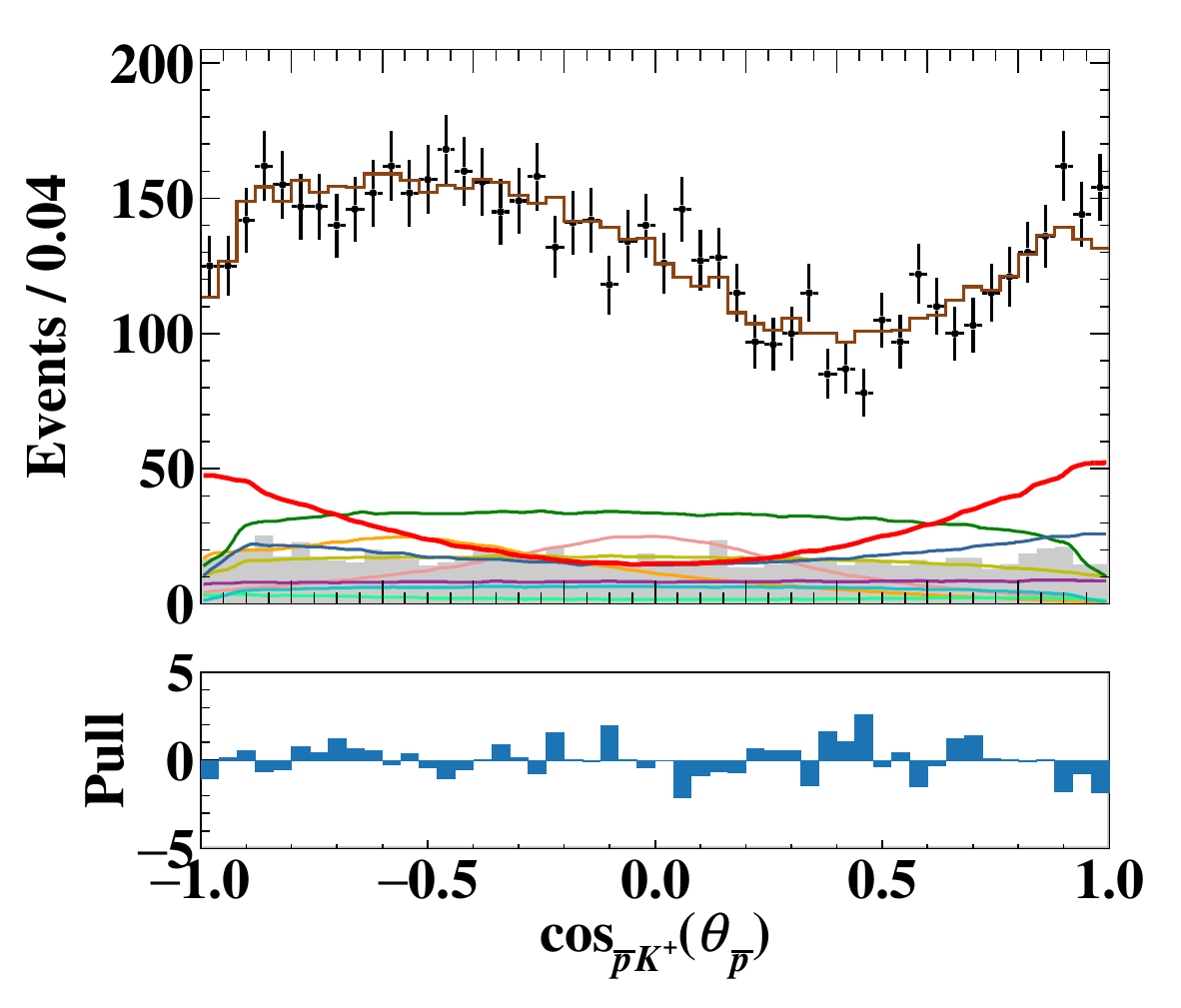}
    \includegraphics[width=0.27\textwidth]{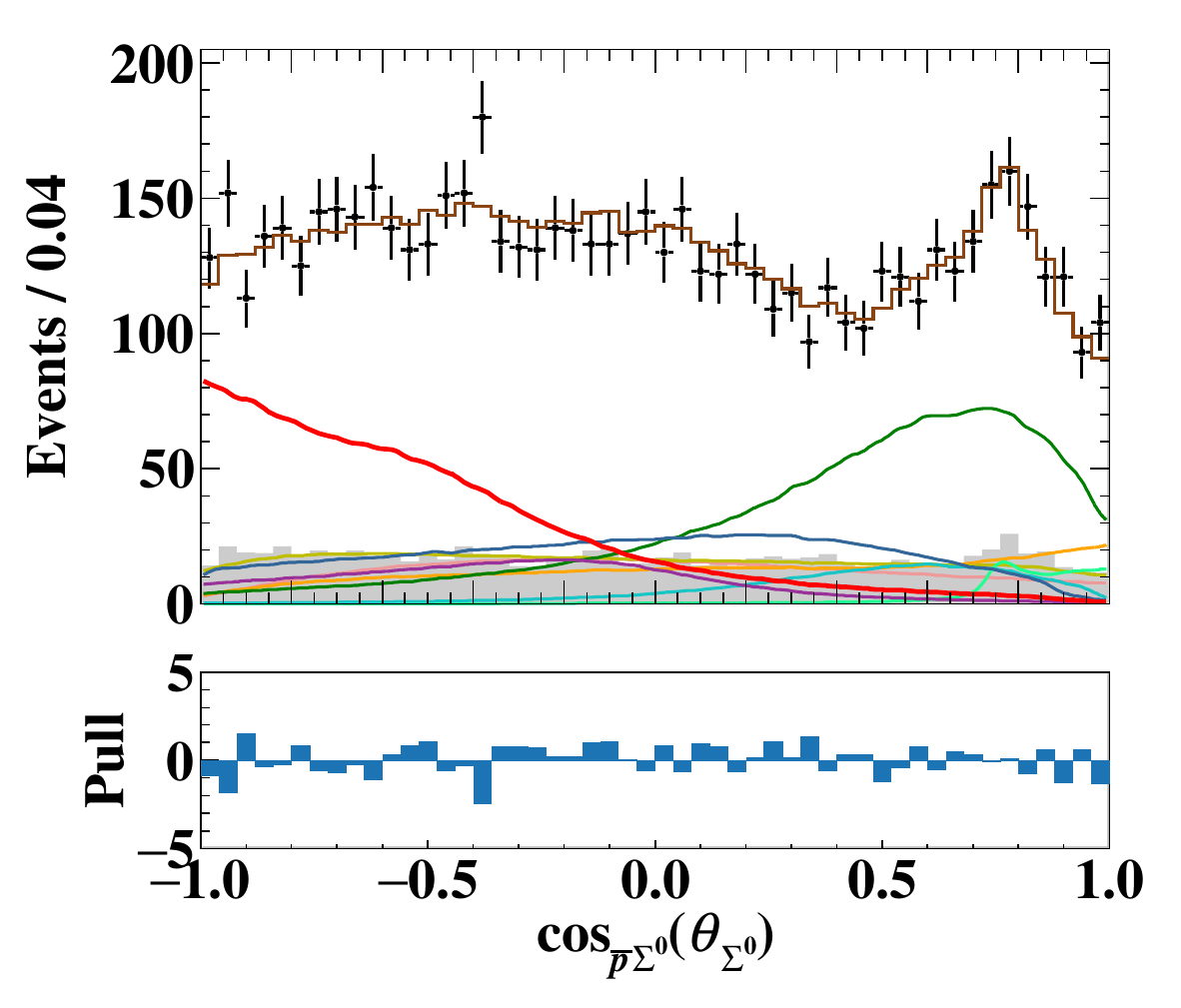}
    \includegraphics[width=0.27\textwidth]{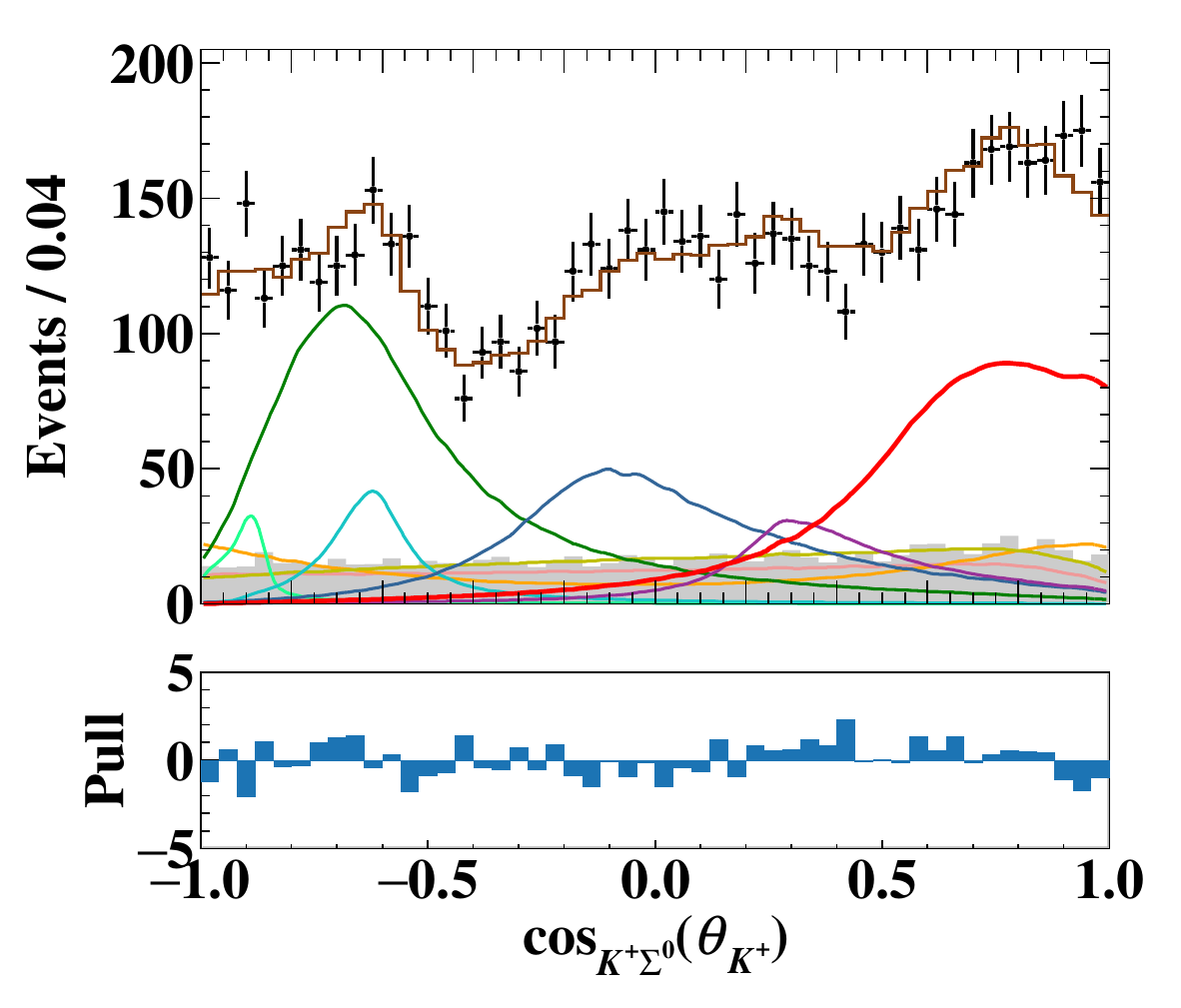}
    \includegraphics[clip, trim=9cm 3.5cm 2.5cm 0cm,width=0.135\textwidth]{Legend.pdf}
    \caption{Distributions of $M_{\bar{p}K^+}$, $M_{\bar{p}\Sigma^0}$, $M_{K^+\Sigma^0}$, $\cos_{\bar{p}K^+}(\theta_{\bar{p}})$, $\cos_{\bar{p}\Sigma^0}(\theta_{\Sigma^0})$ and $\cos_{K^+\Sigma^0}(\theta_{K^+})$.}
    \label{fig:pwa_mass}
\end{figure*}

The nominal fit hypothesis is established by evaluating the statistical significance of each potential component from the change in negative log-likelihood ($\Delta NLL$) and the number of additional degrees of freedom ($\Delta N_\mathrm{dof}$). Components considered are established $N^*$, $\Lambda^*$ and $\Sigma^*$ states with PDG status at least three-star and spin $\le 5/2$~\cite{ParticleDataGroup:2024cfk}, together with $K_2(2250)$, $K_3(2320)$ and an $S$-wave non-resonant term ($NR_{1^-}$) in $M_{\bar{p}\Sigma^0}$.  In addition, four further states---$N(2300)$, $N(2570)$, $\Sigma(2010)$ and $\Sigma(2110)$---are tested in $M_{K^+\Sigma^0}$ and $M_{\bar{p}K^+}$. To describe a prominent structure near $2.33\;\mathrm{GeV}/c^2$ in $M_{\bar{p}K^+}$, we introduce an additional resonance denoted $\Sigma(2330)$. The resonances $N(2300)$, $N(2570)$, $\Lambda(1520)$, $\Sigma(1660)$, $\Sigma(1670)$, $\Sigma(2010)$, $\Sigma(2110)$, $\Sigma(2330)$ and $NR_{1^-}$ are found to be with statistical significance greater than $5\sigma$; no other component exceeds this threshold.  The mass resolution for $M_{\bar{p}K^+}$ is estimated from MC simulation to be $1.2\;\mathrm{MeV}/c^2$, negligible compared with the $\Lambda(1520)$ width of $(15.73\pm0.26)\;\mathrm{MeV}$~\cite{ParticleDataGroup:2024cfk}. Therefore, the $\Lambda(1520)$ mass and width are fixed to the PDG values, while those of the other resonances are left free. Dalitz plots of $M^2_{\bar{p}K^+}$ versus $M^2_{K^+\Sigma^0}$ for data and the normalized PWA result are shown in Fig.~\ref{fig:pwa_dalitz} with the kinematic boundary, and invariant-mass and helicity-angle distributions are given in Fig.~\ref{fig:pwa_mass}. Table~\ref{tab:pwa_result} summarizes the fitted masses, widths and branching fractions; the first uncertainty is statistical and the second systematic.

In the study of $\psi(3686)\to p\bar{p}\pi^0$~\cite{BESIII:2012ssm}, $N(2300)$ and $N(2570)$ were observed for the first time. Here, their spin-parities are confirmed by comparing the likelihood of the nominal hypothesis with those of alternative $J^P$ assignments; the data favor $1/2^+$ for $N(2300)$ and $5/2^-$ for $N(2570)$, consistent with the PDG~\cite{ParticleDataGroup:2024cfk}. Likewise, the one-star states $\Sigma(2010)$ and $\Sigma(2110)$ are tested, and the preferred assignments, $3/2^-$ and $1/2^-$, respectively, also agree with the PDG~\cite{ParticleDataGroup:2024cfk}.

Spin-parity hypotheses for the additional $\Sigma(2330)$ resonance are compared in Table~\ref{tab:JP_scan_S2330}. The $J^P=3/2^-$ assignment yields the highest significance, $11.9\sigma$, relative to a solution that does not include any of the $\Sigma(2330)$ hypotheses. Under this hypothesis the mass, width, and fit fraction (FF) of $\Sigma(2330)$ are measured to be $(2331.9\pm7.6)\;\mathrm{MeV}/c^2$, $(205.7\pm9.5)\;\mathrm{MeV}$, and $(19.8\pm3.4)\%$, respectively, where the uncertainties are statistical only. No established resonance in the PDG~\cite{ParticleDataGroup:2024cfk} matches both the mass and width of $\Sigma(2330)$. A recent theoretical calculation~\cite{Menapara:2024wpb} assigns excited $\Sigma$ states near $2.33\;\mathrm{GeV}/c^2$ to the 1F family; our result is compatible with the predicted 1F$(3/2^-)$ state within $5\;\mathrm{MeV}/c^2$.

\begin{table}[!htbp]
    \centering
    \caption{Statistical significance of different spin-parity hypotheses for the $\Sigma(2330)$ resonance relative to a solution excluding all $\Sigma(2330)$ hypotheses.}
    \label{tab:JP_scan_S2330}
    \begin{tabular}{c@{\qquad\qquad\quad\ }c@{\qquad\qquad\quad\ }c@{\qquad\qquad\quad\ }r}
        \midrule
        \midrule
        $J^P$   & $\Delta NLL$ & $\Delta N_\mathrm{dof}$ & S ($\sigma$) \\
        \midrule
        $1/2^-$ & $20.1$       & $6$                     & $5.1$                   \\
        $1/2^+$ & $48.9$       & $6$                     & $8.9$                   \\
        $3/2^-$ & $85.7$       & $8$                     & $11.9$                  \\
        $3/2^+$ & $71.2$       & $8$                     & $10.7$                  \\
        $5/2^-$ & $74.1$       & $8$                     & $11.0$                  \\
        $5/2^+$ & $53.7$       & $8$                     & $9.1$                   \\
        $7/2^-$ & $44.3$       & $8$                     & $8.0$                   \\
        \midrule
        \midrule
    \end{tabular}
\end{table}

The systematic uncertainties on the branching fraction measurement are split into different categories: reconstruction and event selection of the signal candidates, and the effects which arise from the final fit procedure. The uncertainty on the $\Lambda$ reconstruction efficiency is taken from Ref.~\cite{BESIII:2017kqw}. The uncertainty due to tracking is determined with the method described in Ref.~\cite{Liu:2025trk}. The uncertainty due to PID is determined using control samples of $\jpsi\to K^*\bar{K}$~\cite{BESIII:2011ysp} and $\jpsi\to p\bar{p}\pi^+\pi^-$~\cite{BESIII:2011wmh}, and the difference between the data and MC simulation is regarded as the uncertainty. The uncertainty of photon detection is investigated through the initial state radiation (ISR) process $e^+e^-\to\gamma_\mathrm{ISR}\mu^+\mu^-$. The uncertainty from the kinematic fit is evaluated by the method described in Ref.~\cite{BESIII:2012mpj}, and the track helix parameters are taken from Refs.~\cite{BESIII:2013nam, BESIII:2019gjc}. The uncertainty due to the suppression of the $\chi_{cJ}$ background is estimated by varying the mass window within $10\;\mathrm{MeV}/c^2$, and the difference of branching fraction is taken as the uncertainty. The uncertainty from the signal shape is evaluated from the difference of the signal yield by replacing the nominal shape with a Crystal Ball function convolved with a Gaussian function. The uncertainty from the background is estimated either by replacing the shape of the non-peaking background in the fit with a second-order Chebyshev polynomial function or by varying the number of the continuum background events by $\pm1\sigma$. The maximum deviation in signal yield is adopted as the conservative estimation. To evaluate the uncertainty from the MC model, an alternative amplitude model is used to generate the signal MC sample, and the difference of efficiency between the nominal and alternative models is regarded as the uncertainty. The uncertainties from the branching fractions of the intermediate decays of $\Lambda\to p\pi^-$ and $\Sigma^0\to\Lambda\gamma$ are cited from the PDG~\cite{ParticleDataGroup:2024cfk}. The uncertainty from the total number of $\psip$ events is quoted from Ref.~\cite{BESIII:2024lks}. The total systematic uncertainty, which is the quadratic sum of individual relative systematic uncertainties, is summarized in Table~\ref{tab:br_syu}.

\begin{table}[htb]
    \centering
    \caption{Relative systematic uncertainties (in percent) for the branching fraction measurement.}
    \label{tab:br_syu}
    \begin{tabular}{l@{\qquad\qquad\quad}c}
        \midrule
        \midrule
        Sources                                    & Value$(\%)$ \\
        \midrule
        $\Lambda$ reconstruction                   & 1.1         \\
        Tracking efficiency of $p$ and $K$ from IP & 0.3         \\
        PID efficiency                             & 2.0         \\
        Photon reconstruction                      & 0.5         \\
        Kinematic fit                              & 0.6         \\
        $\chi_{cJ}$ mass window                    & 0.3         \\
        Signal shape                               & 0.6         \\
        Background                                 & 1.8         \\
        MC model                                   & 0.7         \\
        $\mathcal{B}(\Sigma^0\to\gamma\Lambda)$    & 0.0         \\
        $\mathcal{B}(\Lambda\to p\pi)$             & 0.8         \\
        Total number of $\psip$ events             & 0.5         \\
        \midrule
        Total                                      & 3.3         \\
        \midrule
        \midrule
    \end{tabular}
\end{table}

Systematic uncertainties on the masses, widths and fit fractions of the resonances extracted from the PWA are evaluated as follows. To assess the influence of additional components, the PWA is repeated with each of the following resonances added one at a time: $K_2(2250)$, $K_3(2320)$, $N(1700)$, $N(1710)$, $N(1720)$, $N(1875)$, $N(1880)$, $N(1895)$, $N(1900)$, $N(2100)$, $N(2120)$, $\Sigma(1750)$, $\Sigma(1910)$. The uncertainty due to background estimation is obtained by varying the background yields within $\pm1\sigma$ of their statistical uncertainties. The uncertainty associated with the orbital angular momentum of the $NR$ component is estimated by changing the $L$ assignment from 0 to 1 or 2. In the Blatt–Weisskopf barrier factor~\cite{Chung:1993da} the radius parameter is taken to be $d=0.73\;\mathrm{fm}\approx3.7\;\mathrm{GeV}^{-1}$ following Ref.~\cite{BESIII:2019dme}; its uncertainty is evaluated by varying $d$ between $1.0$ and $5.0\;\mathrm{GeV}^{-1}$~\cite{ParticleDataGroup:2024cfk}. The largest deviation observed for each source is taken as the corresponding systematic uncertainty; all sources are assumed to be independent and are added in quadrature. Finally, the systematic uncertainty on the fit fractions is combined with that on the branching fraction quoted above.

In summary, using $(2712.4\pm14.3)\times10^6\ \psi(3686)$ events collected with the BESIII detector, we have performed a partial-wave analysis of $\psi(3686)\to\bar{p}K^+\Sigma^0$ and observed a new excited $\Sigma$ state, $\Sigma(2330)$, in the $M_{\bar{p}K^+}$ distribution with a statistical significance of $11.9\sigma$. The mass and width of $\Sigma(2330)$ are measured to be $(2334.7\pm7.9\pm16.0)\;\mathrm{MeV}/c^2$ and $(206.3\pm9.5\pm18.4)\;\mathrm{MeV}$, respectively, where the first uncertainty is statistical and the second systematic. The spin-parity of $\Sigma(2330)$ favors $3/2^-$; no established resonance in the PDG~\cite{ParticleDataGroup:2024cfk} matches both its mass and width. A recent theoretical prediction~\cite{Menapara:2024wpb} assigns excited $\Sigma$ states near $2.33\;\mathrm{GeV}/c^2$ to the 1F family; our measurement is compatible with the predicted 1F$(3/2^-)$ state within $5\;\mathrm{MeV}/c^2$. The alternative spin-parity hypotheses $3/2^+$ and $5/2^-$ for $\Sigma(2330)$ yield significances of $10.7\sigma$ and $11.0\sigma$, respectively—very close to the nominal $3/2^-$ value. Because of limited statistics and the proximity of the $\Sigma(2330)$ mass to phase-space threshold, the present data cannot conclusively distinguish between these possibilities; further studies with larger data samples and additional decay channels will be required to firmly establish the quantum numbers of $\Sigma(2330)$.
In addition to well-established $\Lambda^*$ and $\Sigma^*$ states such as $\Lambda(1520)$, $\Sigma(1660)$, and $\Sigma(1670)$, the PWA shows that $N(2300)$, $N(2570)$, $\Sigma(2010)$, and $\Sigma(2110)$ are required to describe the data. Although $N(2300)$ and $N(2570)$ are listed as two-star states in the PDG~\cite{ParticleDataGroup:2024cfk}, they are among the dominant contributions to $\psi(3686)\to\bar{p}K^+\Sigma^0$, providing additional evidence for their existence. Spin-parity tests favor $J^P=1/2^+$ for $N(2300)$ and $J^P=5/2^-$ for $N(2570)$, consistent with the assignment by PDG~\cite{ParticleDataGroup:2024cfk}. For the one-star states $\Sigma(2010)$ and $\Sigma(2110)$, the data favor $J^P=3/2^-$ and $J^P=1/2^-$, respectively, again in agreement with the PDG~\cite{ParticleDataGroup:2024cfk}.  
The branching fraction of $\psi(3686)\to\bar{p}K^+\Sigma^0$ is measured to be either $(2.44\pm0.20\pm0.08)\times10^{-5}$ or $(1.73\pm0.29\pm0.06)\times10^{-5}$, where the first uncertainty is statistical and the second systematic; the two solutions arise from an ambiguous phase between resonant and continuum amplitudes. 

\begin{table*}[htb]
    \centering
    \caption{Masses, widths, and production branching fractions of the resonances in the PWA fit using the nominal fit hypothesis of amplitudes. The first uncertainties are statistical and the second systematic.}
    \label{tab:pwa_result}
    \begin{tabular}{c@{\qquad\qquad\qquad\qquad}r@{$\;\pm\;$}r@{$\;\pm\;$}r@{\qquad\qquad\qquad\qquad}r@{$\;\pm\;$}r@{$\;\pm\;$}r@{\qquad\qquad\qquad\qquad}r@{$\;\pm\;$}c@{$\;\pm\;$}l}
        \midrule
        \midrule
        Resonance      & \multicolumn{3}{c@{\qquad\qquad\qquad\qquad}}{Mass $(\mathrm{MeV}/c^2)$} & \multicolumn{3}{c@{\qquad\qquad\qquad\qquad}}{Width ($\mathrm{MeV}$)} & \multicolumn{3}{c}{$\mathcal{B}\;(\times10^{-6})$}                                                        \\
        \midrule
        $N(2300)$       & $2285.3$                                       & $13.8$                                      & $26.1$                                             & $335.9$ & $14.6$ & $21.6$ & $1.95$ & $0.53$ & $0.58$ \\
        $N(2570)$       & $2577.9$                                       & $14.8$                                      & $33.2$                                             & $255.1$ & $14.6$ & $19.7$ & $2.28$ & $0.53$ & $0.67$ \\
        $\Lambda(1520)$ & \multicolumn{3}{c@{\qquad\qquad\qquad\qquad}}{---}                       & \multicolumn{3}{c@{\qquad\qquad\qquad\qquad}}{---}                    & $0.49$                                             & $0.10$  & $0.14$                                     \\
        $\Sigma(1660)$  & $1680.5$                                       & $13.8$                                      & $21.3$                                             & $247.7$ & $15.8$ & $18.8$ & $5.84$ & $0.69$ & $1.55$ \\
        $\Sigma(1670)$  & $1680.5$                                       & $4.1$                                       & $11.0$                                             & $73.4$  & $9.1$  & $7.1$  & $1.22$ & $0.38$ & $0.28$ \\
        $\Sigma(2010)$  & $1980.1$                                       & $10.0$                                      & $18.1$                                             & $192.5$ & $13.3$ & $13.3$ & $3.11$ & $0.50$ & $0.92$ \\
        $\Sigma(2110)$  & $2172.4$                                       & $6.3$                                       & $10.9$                                             & $107.2$ & $9.3$  & $10.2$ & $1.17$ & $0.23$ & $0.38$ \\
        $\Sigma(2330)$  & $2334.7$                                       & $7.9$                                       & $16.0$                                             & $206.3$ & $9.5$  & $18.4$ & $4.47$ & $0.58$ & $1.52$ \\
        \midrule
        \midrule
    \end{tabular}
\end{table*}

\begin{acknowledgments}
{\it Acknowledgments—}
The BESIII Collaboration thanks the staff of BEPCII (https://cstr.cn/31109.02.BEPC) and the IHEP computing center for their strong support. This work is supported in part by National Key R\&D Program of China under Contracts Nos. 
2025YFA1613900, 2023YFA1606000, 2023YFA1606704; 
National Natural Science Foundation of China (NSFC) under Contracts Nos. 
12247101, 11635010, 11935015, 11935016, 11935018, 12025502, 12035009, 12035013, 12061131003, 12192260, 12192261, 12192262, 12192263, 12192264, 12192265, 12221005, 12225509, 12235017, 12361141819; 
the Fundamental Research Funds for the Central Universities No. lzujbky-2025-ytA05,  No. lzujbky-2025-it06,  No. lzujbky-2024-jdzx06;
the Natural Science Foundation of Gansu Province No. 22JR5RA389, No.25JRRA799;
the ‘111 Center’ under Grant No. B20063;
the Chinese Academy of Sciences (CAS) Large-Scale Scientific Facility Program; the Strategic Priority Research Program of Chinese Academy of Sciences under Contract No. XDA0480600; CAS under Contract No. YSBR-101; 100 Talents Program of CAS; The Institute of Nuclear and Particle Physics (INPAC) and Shanghai Key Laboratory for Particle Physics and Cosmology; ERC under Contract No. 758462; German Research Foundation DFG under Contract No. FOR5327; Istituto Nazionale di Fisica Nucleare, Italy; Knut and Alice Wallenberg Foundation under Contracts Nos. 2021.0174, 2021.0299; Ministry of Development of Turkey under Contract No. DPT2006K-120470; National Research Foundation of Korea under Contract No. NRF-2022R1A2C1092335; National Science and Technology fund of Mongolia; Polish National Science Centre under Contract No. 2024/53/B/ST2/00975; STFC (United Kingdom); Swedish Research Council under Contract No. 2019.04595; U. S. Department of Energy under Contract No. DE-FG02-05ER41374.
\end{acknowledgments}

\bibliography{ref.bib}

@article{Edwards:2011jj,
  author        = {Edwards, Robert G. and Dudek, Jozef J. and Richards, David G. and Wallace, Stephen J.},
  title         = {{Excited state baryon spectroscopy from lattice QCD}},
  doi           = {10.1103/PhysRevD.84.074508},
  journal       = {Phys. Rev. D},
  volume        = {84},
  pages         = {074508},
  year          = {2011}
}

@article{Klempt:2009pi,
  author        = {Klempt, Eberhard and Richard, Jean-Marc},
  title         = {{Baryon spectroscopy}},
  doi           = {10.1103/RevModPhys.82.1095},
  journal       = {Rev. Mod. Phys.},
  volume        = {82},
  pages         = {1095--1153},
  year          = {2010}
}

@article{BESIII:2024lks,
  author        = {Ablikim, Medina and others},
  collaboration = {BESIII},
  title         = {{Determination of the number of $\psi(3686)$ events taken at BESIII}},
  doi           = {10.1088/1674-1137/ad595b},
  journal       = {Chin. Phys. C},
  volume        = {48},
  number        = {9},
  pages         = {093001},
  year          = {2024}
}

@article{BESIII:2009fln,
  author        = {Ablikim, M. and others},
  collaboration = {BESIII},
  title         = {{Design and Construction of the BESIII Detector}},
  doi           = {10.1016/j.nima.2009.12.050},
  journal       = {Nucl. Instrum. Meth. A},
  volume        = {614},
  pages         = {345--399},
  year          = {2010}
}

@inproceedings{Yu:2016cof,
  author    = {Yu, Chenghui and others},
  title     = {{BEPCII Performance and Beam Dynamics Studies on Luminosity}},
  booktitle = {{7th International Particle Accelerator Conference}},
  doi       = {10.18429/JACoW-IPAC2016-TUYA01},
  pages     = {TUYA01},
  year      = {2016}
}

@article{BESIII:2020nme,
  author        = {Ablikim, M. and others},
  collaboration = {BESIII},
  title         = {{Future Physics Programme of BESIII}},
  doi           = {10.1088/1674-1137/44/4/040001},
  journal       = {Chin. Phys. C},
  volume        = {44},
  number        = {4},
  pages         = {040001},
  year          = {2020}
}

@article{Zhang:2022bdc,
  author  = {Zhang, Jia-Wei and others},
  title   = {{Suppression of top-up injection backgrounds with offline event filter in the BESIII experiment}},
  doi     = {10.1007/s41605-022-00331-7},
  journal = {Radiat. Detect. Technol. Methods},
  volume  = {6},
  number  = {3},
  pages   = {289--293},
  year    = {2022}
}

@article{GEANT4:2002zbu,
  author        = {Agostinelli, S. and others},
  collaboration = {GEANT4},
  title         = {{GEANT4--a simulation toolkit}},
  doi           = {10.1016/S0168-9002(03)01368-8},
  journal       = {Nucl. Instrum. Meth. A},
  volume        = {506},
  pages         = {250--303},
  year          = {2003}
}

@article{Jadach:2000ir,
  author        = {Jadach, S. and Ward, B. F. L. and Was, Z.},
  title         = {{Coherent exclusive exponentiation for precision Monte Carlo calculations}},
  doi           = {10.1103/PhysRevD.63.113009},
  journal       = {Phys. Rev. D},
  volume        = {63},
  pages         = {113009},
  year          = {2001}
}

@article{Jadach:1999vf,
  author        = {Jadach, S. and Ward, B. F. L. and Was, Z.},
  title         = {{The Precision Monte Carlo event generator $KK$ for two fermion final states in $e^+e^-$ collisions}},
  doi           = {10.1016/S0010-4655(00)00048-5},
  journal       = {Comput. Phys. Commun.},
  volume        = {130},
  pages         = {260--325},
  year          = {2000}
}

@article{Lange:2001uf,
  author  = {Lange, D. J.},
  editor  = {Erhan, S. and Schlein, P. and Rozen, Y.},
  title   = {{The EvtGen particle decay simulation package}},
  doi     = {10.1016/S0168-9002(01)00089-4},
  journal = {Nucl. Instrum. Meth. A},
  volume  = {462},
  pages   = {152--155},
  year    = {2001}
}

@article{Ping:2008zz,
  author  = {Ping, Rong-Gang},
  title   = {{Event generators at BESIII}},
  doi     = {10.1088/1674-1137/32/8/001},
  journal = {Chin. Phys. C},
  volume  = {32},
  pages   = {599},
  year    = {2008}
}

@article{ParticleDataGroup:2024cfk,
  author        = {Navas, S. and others},
  collaboration = {Particle Data Group},
  title         = {{Review of particle physics}},
  doi           = {10.1103/PhysRevD.110.030001},
  journal       = {Phys. Rev. D},
  volume        = {110},
  number        = {3},
  pages         = {030001},
  year          = {2024}
}

@article{Chen:2000tv,
  author  = {Chen, J. C. and Huang, G. S. and Qi, X. R. and Zhang, D. H. and Zhu, Y. S.},
  title   = {{Event generator for $J/\psi$ and $\psi (2S)$ decay}},
  doi     = {10.1103/PhysRevD.62.034003},
  journal = {Phys. Rev. D},
  volume  = {62},
  pages   = {034003},
  year    = {2000}
}

@article{Zhou:2020ksj,
  author        = {Zhou, Xingyu and Du, Shuxian and Li, Gang and Shen, Chengping},
  title         = {{TopoAna: A generic tool for the event type analysis of inclusive Monte-Carlo samples in high energy physics experiments}},
  doi           = {10.1016/j.cpc.2020.107540},
  journal       = {Comput. Phys. Commun.},
  volume        = {258},
  pages         = {107540},
  year          = {2021}
}

@article{Chung:1993da,
  author  = {Chung, S. U.},
  title   = {{Helicity coupling amplitudes in tensor formalism}},
  doi     = {10.1103/PhysRevD.48.1225},
  journal = {Phys. Rev. D},
  volume  = {48},
  pages   = {1225--1239},
  year    = {1993},
}

@article{BESIII:2019dme,
  author        = {Ablikim, M. and others},
  collaboration = {BESIII},
  title         = {{Partial wave analysis of $\psi(3686)\to K^{+}K^{-}\eta$}},
  doi           = {10.1103/PhysRevD.101.032008},
  journal       = {Phys. Rev. D},
  volume        = {101},
  number        = {3},
  pages         = {032008},
  year          = {2020}
}

@article{BESIII:2012ssm,
  author        = {Ablikim, M. and others},
  collaboration = {BESIII},
  title         = {{Observation of two new $N^*$ resonances in the decay $\psi(3686)\to p\bar{p}\pi^0$}},
  doi           = {10.1103/PhysRevLett.110.022001},
  journal       = {Phys. Rev. Lett.},
  volume        = {110},
  number        = {2},
  pages         = {022001},
  year          = {2013}
}

@article{Guo:2022gkg,
  author        = {Guo, Y. P. and Yuan, C. Z.},
  title         = {{Impact of the interference between the resonance and continuum amplitudes on vector quarkonia decay branching fraction measurements}},
  doi           = {10.1103/PhysRevD.105.114001},
  journal       = {Phys. Rev. D},
  volume        = {105},
  number        = {11},
  pages         = {114001},
  year          = {2022}
}

@article{BESIII:2017kqw,
  author        = {Ablikim, Medina and others},
  collaboration = {BESIII},
  title         = {{Study of $J/\psi$ and $\psi(3686)$ decay to $\Lambda\bar{\Lambda}$ and $\Sigma^0\bar{\Sigma}^0$ final states}},
  doi           = {10.1103/PhysRevD.95.052003},
  journal       = {Phys. Rev. D},
  volume        = {95},
  number        = {5},
  pages         = {052003},
  year          = {2017}
}

@article{BESIII:2011ysp,
  author        = {Ablikim, M. and others},
  collaboration = {BESIII},
  title         = {{Study of $\chi_{cJ}$ radiative decays into a vector meson}},
  doi           = {10.1103/PhysRevD.83.112005},
  journal       = {Phys. Rev. D},
  volume        = {83},
  pages         = {112005},
  year          = {2011}
}

@article{BESIII:2011wmh,
  author        = {Ablikim, M. and others},
  collaboration = {BESIII},
  title         = {{Search for a light exotic particle in $J/\psi$ radiative decays}},
  doi           = {10.1103/PhysRevD.85.092012},
  journal       = {Phys. Rev. D},
  volume        = {85},
  pages         = {092012},
  year          = {2012}
}

@article{BESIII:2013nam,
  author        = {Ablikim, M. and others},
  collaboration = {BESIII},
  title         = {{Search for $\eta_c(2S)h_c\to p\bar{p}$ decays and measurements of the $\chi_{cJ}\to p\bar{p}$ branching fractions}},
  doi           = {10.1103/PhysRevD.88.112001},
  journal       = {Phys. Rev. D},
  volume        = {88},
  number        = {11},
  pages         = {112001},
  year          = {2013}
}

@article{BESIII:2019gjc,
  author        = {Ablikim, M. and others},
  collaboration = {BESIII},
  title         = {{Cross section measurements of $e^+ e^-\to\omega\chi_{c0}$ form $\sqrt{s}=$ 4.178 to 4.278 GeV}},
  doi           = {10.1103/PhysRevD.99.091103},
  journal       = {Phys. Rev. D},
  volume        = {99},
  number        = {9},
  pages         = {091103},
  year          = {2019}
}

@article{BESIII:2012mpj,
  author        = {Ablikim, M. and others},
  collaboration = {BESIII},
  title         = {{Search for hadronic transition $\chi_{cJ}\to\eta_c\pi^+\pi^-$ and observation of $\chi_{cJ}\to K\bar{K}\pi\pi\pi$}},
  doi           = {10.1103/PhysRevD.87.012002},
  journal       = {Phys. Rev. D},
  volume        = {87},
  number        = {1},
  pages         = {012002},
  year          = {2013}
}

@article{BESIII:2012koo,
  author        = {Ablikim, M. and others},
  collaboration = {BESIII},
  title         = {{Measurements of $\psi^\prime \to \bar{p} K^+ \Sigma^0$ and $\chi_{cJ} \to \bar{p} K^+ \Lambda$}},
  doi           = {10.1103/PhysRevD.87.012007},
  journal       = {Phys. Rev. D},
  volume        = {87},
  number        = {1},
  pages         = {012007},
  year          = {2013}
}

@article{BESIII:2022cxi,
  author        = {Ablikim, M. and others},
  collaboration = {BESIII},
  title         = {{Measurement of $\psi(3686)\to\Lambda\bar{\Lambda}\eta$ and $\psi(3686)\to\Lambda\bar{\Lambda}\pi^0$ decays}},
  doi           = {10.1103/PhysRevD.106.072006},
  journal       = {Phys. Rev. D},
  volume        = {106},
  number        = {7},
  pages         = {072006},
  year          = {2022}
}

@article{Chung:1971ri,
  author       = {Chung, Suh Urk},
  title        = {{SPIN FORMALISMS}},
  journal      = {},
  reportnumber = {CERN-71-08},
  doi          = {10.5170/CERN-1971-008},
  month        = {3},
  year         = {1971}
}

@article{Richman:1984gh,
  author       = {Richman, Jeffrey D.},
  title        = {{An Experimenter's Guide to the Helicity Formalism}},
  journal      = {},
  reportnumber = {CALT-68-1148},
  month        = {6},
  year         = {1984}
}

@misc{Jiang:2020tf,
  author       = {Y.~Jiang and others},
  howpublished = {\url{https://github.com/jiangyi15/tf-pwa}},
  title        = {{Open-source framework TF-PWA package}},
  year         = {2020}
}

@article{Menapara:2024wpb,
  author        = {Menapara, Chandni and Rai, Ajay Kumar},
  title         = {{Hadron Spectroscopy: Light, Strange Baryons}},
  doi           = {10.1007/s00601-024-01933-1},
  journal       = {Few Body Syst.},
  volume        = {65},
  number        = {2},
  pages         = {63},
  year          = {2024}
}

@article{BESIII:2024dmr,
  author        = {Ablikim, Medina and others},
  collaboration = {BESIII},
  title         = {{Measurement of $\Sigma^+$ transverse polarization in $e^+e^-$ collisions at $ \sqrt{s} $ = 3.68 \ensuremath{-} 3.71 GeV}},
  doi           = {10.1007/JHEP12(2024)186},
  journal       = {J. High Energy Phys.},
  volume        = {12},
  pages         = {186 (2024)},
  year          = {2024}
}

@article{Liu:2025trk,
  author   = {Liu, Fang and others},
  title    = {Study of the tracking efficiency of charged pions at BESIII},
  journal  = {Radiat. Detect. Technol. Methods},
  year     = {2025},
  month    = {Feb},
  day      = {05},
  issn     = {2509-9949},
  doi      = {10.1007/s41605-025-00530-y}
}

@article{BESIII:2022fhe,
  author        = {Ablikim, Medina and others},
  collaboration = {BESIII},
  title         = {{Study of $\psi(3686)\to\Lambda\bar{\Lambda}\omega$}},
  doi           = {10.1103/PhysRevD.106.112011},
  journal       = {Phys. Rev. D},
  volume        = {106},
  number        = {11},
  pages         = {112011},
  year          = {2022}
}

@article{BESIII:2023syz,
  author        = {Ablikim, M. and others},
  collaboration = {BESIII},
  title         = {{Precise measurement of the branching fractions of $J/\psi\to\bar{\Lambda}\pi^+\Sigma^-$+c.c. and $J/\psi\to\bar{\Lambda}\pi^-\Sigma^+$+c.c.}},
  doi           = {10.1103/PhysRevD.108.112012},
  journal       = {Phys. Rev. D},
  volume        = {108},
  number        = {11},
  pages         = {112012},
  year          = {2023}
}

@article{BESIII:2024vqu,
  author        = {Ablikim, Medina and others},
  collaboration = {BESIII},
  title         = {{Partial wave analyses of $\psi(3686)\to p\bar{p}\pi^0$ and $\psi(3686)\to p\bar{p}\eta$}},
  doi           = {10.1103/PhysRevD.111.032011},
  journal       = {Phys. Rev. D},
  volume        = {111},
  number        = {3},
  pages         = {032011},
  year          = {2025}
}

@article{BESIII:2024jgy,
  author        = {Ablikim, Medina and others},
  collaboration = {BESIII},
  title         = {{Partial wave analysis of $ \psi (3686)\to \Lambda {\overline{\Sigma}}^0{\pi}^0 $ + c.c}},
  doi           = {10.1007/JHEP02(2025)212},
  journal       = {J. High Energy Phys.},
  volume        = {02},
  pages         = {212 (2025)},
  year          = {2025}
}

@article{BESIII:2021cxx,
    author = "Ablikim, M. and others",
    collaboration = "BESIII",
    title = "{Number of $J/\psi$ events at BESIII}",
    doi = "10.1088/1674-1137/ac5c2e",
    journal = "Chin. Phys. C",
    volume = "46",
    number = "7",
    pages = "074001",
    year = "2022"
}

@article{BESIII:2023euh,
    author = "Ablikim, Medina and others",
    collaboration = "BESIII",
    title = "{Measurement of $\bar\Lambda$ transverse polarization in $e^{+}e^{-}$ collisions at $\sqrt{s} $ = 3.68- 3.71 GeV}",
    eprint = "2303.00271",
    archivePrefix = "arXiv",
    primaryClass = "hep-ex",
    doi = "10.1007/JHEP10(2023)081",
    journal = "JHEP",
    volume = "10",
    pages = "081",
    year = "2023",
    note = "[Erratum: JHEP 12, 080 (2023)]"
}

\begin{widetext}
\centering
M.~Ablikim$^{1}$\BESIIIorcid{0000-0002-3935-619X},
M.~N.~Achasov$^{4,b}$\BESIIIorcid{0000-0002-9400-8622},
P.~Adlarson$^{81}$\BESIIIorcid{0000-0001-6280-3851},
X.~C.~Ai$^{86}$\BESIIIorcid{0000-0003-3856-2415},
R.~Aliberti$^{39}$\BESIIIorcid{0000-0003-3500-4012},
A.~Amoroso$^{80A,80C}$\BESIIIorcid{0000-0002-3095-8610},
Q.~An$^{77,64,\dagger}$,
Y.~Bai$^{62}$\BESIIIorcid{0000-0001-6593-5665},
O.~Bakina$^{40}$\BESIIIorcid{0009-0005-0719-7461},
Y.~Ban$^{50,g}$\BESIIIorcid{0000-0002-1912-0374}
% ...
\\
\vspace{0.2cm}
(BESIII Collaboration)\\
\vspace{0.2cm} {\it
$^{1}$ Institute of High Energy Physics, Beijing 100049, People's Republic of China\\
% ...
$^{4}$ Budker Institute of Nuclear Physics SB RAS (BINP), Novosibirsk 630090, Russia\\
% ...
$^{39}$ Johannes Gutenberg University of Mainz, Johann-Joachim-Becher-Weg 45, D-55099 Mainz, Germany\\
$^{40}$ Joint Institute for Nuclear Research, 141980 Dubna, Moscow region, Russia\\
% ...
$^{50}$ Peking University, Beijing 100871, People's Republic of China\\
% ...
$^{62}$ Southeast University, Nanjing 211100, People's Republic of China\\
% ...
$^{64}$ State Key Laboratory of Particle Detection and Electronics, Beijing 100049, Hefei 230026, People's Republic of China\\
% ...
$^{77}$ University of Science and Technology of China, Hefei 230026, People's Republic of China\\
% ...
$^{80}$ University of Turin and INFN, (A)University of Turin, I-10125, Turin, Italy; (B)University of Eastern Piedmont, I-15121, Alessandria, Italy; (C)INFN, I-10125, Turin, Italy\\
$^{81}$ Uppsala University, Box 516, SE-75120 Uppsala, Sweden\\
% ...
$^{86}$ Zhengzhou University, Zhengzhou 450001, People's Republic of China\\

\vspace{0.2cm}
$^{\dagger}$ Deceased\\
% ...
$^{b}$ Also at the Novosibirsk State University, Novosibirsk, 630090, Russia\\
% ...
$^{g}$ Also at State Key Laboratory of Nuclear Physics and Technology, Peking University, Beijing 100871, People's Republic of China\\
% ...
}

\end{widetext}
\end{document}